IAC-22-C2.6.x68465

# Pourable and Destroyable Cosmic Ray Radiation Shield for Spacecraft

**Jarred C Novak**

*Physic Department, Wichita State University, 1845 Fairmount St., Wichita, Kansas, United States 67212, jcnovak@shockers.wichita.edu*

**Abstract**

Historically, materials such as lead, tungsten, and iron have been used in spacecraft to shield scientific detectors from Cosmic Rays. These materials work well when re-entry to Earth is not an issue. The typical strategy is to have a controlled descent of the spacecraft or to have extremely limited shielding, if any, due to NASA's requirement that all impacting parts must impact with no greater than 15J of energy. Given the nature of this mission neither a controlled descent nor having no shielding was not an option. This is the issue Wichita State University (WSU) nuSOL (Neutrino Solar Orbiting Laboratory) team is facing for its 3U CubeSat demonstrator.

The CubeSat will be equipped with scientific equipment with the purpose of detecting solar neutrinos, and the less background noise from Cosmic Rays the better the study will be. Through simulations, density tests, and burn tests, WSU was able to develop an epoxy-based shield doped with either iron or tungsten powder. The simulations were conducted by firing electrons, protons, alpha particles, and oxygen and iron nuclei into the shield material with energies ranging from 1MeV until consistent failure rate using Geant4 (Geometry and Tracking). The standards for these simulations are the base epoxy at $1.15g/cm^3$ to solid steel at $8g/cm^3$. Mixing tests have determined for iron, a density of $4g/cm^3$ is achievable, which is 53% iron by volume. tungsten epoxy with a density of $7.5g/cm^3$ is more easily achieved, and results in 40% tungsten by volume. These ratios are concrete in texture, pourable, and homogeneous.

With the data collected, several prototypes of varying densities were made by pouring the mixture into molds. The results indicate that the $4g/cm^3$ iron doped epoxy does not have a 90% punch-through until 40MeV for electrons and 74MeV for protons. The tungsten fared better at 1GeV and 90MeV respectively, with the proton never exceeding 94% punch-through at 1GeV. The steel 90% fail are 250MeV for electron and 125MeV for proton. Due to the current requirements of the mission, the densities of iron $4g/cm^3$ and tungsten $7.5g/cm^3$ have been determined to be the best fit for the nuSOL project. Both materials have been tested to determine if they burn upon re-entry and neither shield survived past 425°C. A vibration test is planned to ensure survival of launch along with a test to measure the science package and shielding capabilities for accelerator particles to ensure efficiency.

**Acronyms/Abbreviations**
Wichita State University (WSU)
Neutrino Solar Orbiting Laboratory (nuSOL)
National Aeronautics and Space Administration (NASA)
Geometry and Tracking (Geant4)
Polyethylene Terephthalate Glycol (PETG)
Computer Aided Design (CAD)

## 1. Introduction

The nuSOL project is developing a space probe that is to detect neutrinos near Sun solar orbit [1]. As with all new experiments it is the responsibility of the experimental team to determine that the new innovative technologies will work. One way to accomplish that is to have test flights of the equipment. The nuSOL team has proven that the technologies work on Earth, so the next step is to accomplish a test space mission. This mission will be conducted by sending the neutrino detector to space in the form of a 3U CubeSat to orbit the Earth for approximately 2 years to insure it collects data correctly and efficiently [2]. To ensure that this happens a radiation shield was needed to block out as many Cosmic Rays as possible to reduce the background noise on the detector.

### 1.1 Major Requirements

Major obstacles for this test mission include how to protect the equipment from high energy particles and ensure NASA requirements are met. While these issues are not new, the process and the approach to the shield design and materials are. Initially the WSU team thought that an solid iron shield would be the best way to achieve this goal. iron has a density of $7.874g/cm^3$ and has an initial melting temperature of 1538°C. The only problem is that anything we send into Earth orbit needs to be able to burn up in the atmosphere when the detector returns to Earth. The average re-entry temperature 1871°C, and while that is above the initial melting temperature for iron, other factors such as the size and speed of the object need to be considered. With the dimensions of the 3U CubeSat and the need of iron shielding to be no less than 1.5cm in thickness for effective shielding, it was determined that upon re-entry that a solid iron shield would not completely melt. It was also determined that there was a high probability of the iron shield falling to Earth at greater than 15J of energy. Since this spacecraft has no re-entry engine or control system for its descent, a different type of shielding material is required. After evaluating different types of materials, it was determined that an epoxy-based shield doped with either tungsten or iron would be the best fit for this mission.

*1.2 Shield Attachment*

Not only was the material and shield design new, but also the way the shield interfaces and attaches to the CubeSat needed to be an innovative design. This process has many moving parts given that multiple organizations make up the nuSOL team and are responsible for individual components. These organizations include WSU, NASA's Jet Propulsion Laboratory and Marshall Space Flight Center, and Nanoavionics. It is the responsibility WSU to design the shield, the shield attachment system, and to incorporate all other components to fit within the 3U CubeSat made by Nanoavionics. The main driving force behind the complete CubeSat design is for WSU to have the ability to attach the shield in a manner compliant with all other mechanical and science packages while keeping the integrity and efficiency of the CubeSat mission. With an innovative design the nuSOL team at WSU believes this has been accomplished, but further tests including a vibration test, G-force test, and stress/strain test are scheduled to confirm this. Also, high energy particle test will be conducted to test the science package and shield efficiency at Fermi National Laboratory and is scheduled for January 2023.

**2. Materials**

The materials for this CubeSat mission are a critical obstacle that needed to be addressed early on. The nuSOL team has created a new science package for neutrino detection, but as with most experiments a test mission must be accomplished to assure the science can be achieved. The CubeSat flight is the test for the final satellite mission that will be headed towards the Sun. Unlike the mission to the Sun, which will have extensive shielding and will not be coming back to Earth, the CubeSat does not have this luxury. So, the project team needed to create a way to shield the detector while still meeting NASA guidelines on returning spacecraft. This led the WSU team to design and create a innovative way to shield the science package from radiation.

*2.1 Shield Material*

For this project we need something that is durable, able to be molded, and can adhere easily. The obvious choice was to use a polyester resin. It was decided to use 3M 2216 translucent epoxy adhesive resin. This resin has resilient material properties to the point where it is often used in equipment when normal mechanical fittings will not suffice or have been broken. The design that we are using for the science package includes specialized and fragile equipment and electronics (Figure 1), so the durability and flexibility of the resin was the driving force.

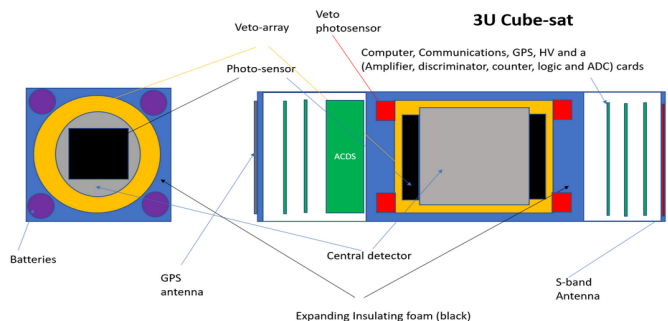

*Figure 1: Block diagram of science package and electronics [2]*

This is a two-part resin that consist of a base and accelerator that we were able to mix in the iron or tungsten filler and with the help of a 3D mold, create the desired shape of a shield prototype. It also helped that the resin has a relatively long work life at 120 minutes. Other material properties that went into determining this resin was the overlap shear test which at lower temperatures was 3000 psi and the shear modulus which ranged from 342Mpa to 2745Mpa when tested at a range of temperatures with the lower end being at higher temperatures, and viscosity of 5000-9000CPS [2]. These properties are important due to the temperatures the CubeSat will encounter in space and with the shielding having to be attached to a metal frame for the ejection out of a rocket to enter orbit and for pourability respectfully.

*2.1.1 iron and tungsten*

For the initial study, iron filler was used to determine the maximum density and ratio achievable. iron was originally determined based on the standard use of solid steel as a radiation shield, and since iron does not become activated with exposure to radiation, it seemed like a promising candidate. Through the process of the experiment elaborated in section 3, it was determined that a density of $4/cm^3$ was obtainable and consisted of a ratio that is 53% iron by volume. This achievement was ground-breaking for the nuSOL project and solidified that it is possible to have some radiation protection during the CubeSat mission. While the iron doped epoxy gives a significant increase in radiation protection, the original goal of the radiation shield design was to get as close to the 8g/cm$^3$ density of steel, it left much to be desired.

After some deliberation, it was decided that tungsten filler should also be studied. With tungsten having a higher density than iron, the ability to create a doped epoxy with similar densities to steel seemed more achievable. This turned out to be true, with samples and prototypes ranging from 3.5g/cm$^3$ to 7.5g/cm$^3$ and a volume ratio of 10% to 40% tungsten. The limitations of a tungsten doped epoxy shield were confined to the payload weight, rather that the structural integrity of the shield itself. The only downfall to using tungsten filler is that after a period of approximately 2 years the metal will become activated with constant radiation exposure. Given that the nuSOL CubeSat mission will collect data and orbit less than this mark it makes a great shield and performs not only better than the iron doped epoxy, but also the solid steel in blocking Cosmic Ray radiation.

*2.2 Shield Attachment*

The final piece was to design a way to attach the tungsten doped epoxy shield to the Nanoavionics provided CubeSat bus. The design itself is expanded upon in section 3.3, but the materials need to be lightweight yet structurally strong. This is to ensure minimal movement of the shield during launch operations while staying under the payload weight limit of 8kg. Aircraft structural Aluminum alloy 2024 was chosen for much of the attachment assembly for these reasons. It is lighter than steel counterparts and still can be fabricated at WSU to the custom design. The other portion of the assembly are steel axle keys to ensure little to no gaps are present between the shield and the CubeSat bus.

## 3. Methods

For this study there were several methods that were used to determine the effectiveness of the shielding and its ability to be created easily. For the CubeSat we need to stop most high energy particles up to the 50 MeV range. This does not mean everything below that level will be stopped but for the detector to work properly and get the most accurate data the vast majority cannot enter the device. To accomplish this, we ran simulations at various densities and energy levels. To ensure that we were able to achieve the corresponding density in real life we needed to design and create experimental test. These tests consisted of mixing ratios to understand what could be achieved, height test to ensure the density was homogeneous, and finally prototypes. This was not only to make sure we could create such a shield but to understand the manufacturing process and create a method to re-create such a shield with high precision.

*3.1 Simulations*

The team at WSU have studied how to properly construct the CubeSat shielding, which was done in the summer of 2021. To determine the effectiveness of both the iron and tungsten doped epoxies, the initial step of this process was to conduct simulations using Geant4. Geant4 is an C++ extension library for particle physics simulation. These simulations consisted of shooting electrons, protons, alpha particles, and oxygen and iron nuclei at 1.5cm thick doped epoxy at energy levels ranging from 1MeV and up until a consistent failure rate was reached. An event was considered a failure if the particle punched through, i.e. if the particle or shower deposited any energy in the volume beyond the shield (Figure 2). The data of interest was the percent of particles that were getting through the shield at these energy levels. We used a steel plate and non-doped epoxy as standards for reference, then changed the doped epoxy density ranging from 1g/cm$^3$ to 5g/cm$^3$ for iron and 3.5g/cm$^3$ to 7.5g/cm$^3$ for tungsten. The results of these simulations are discussed in section 4.1.

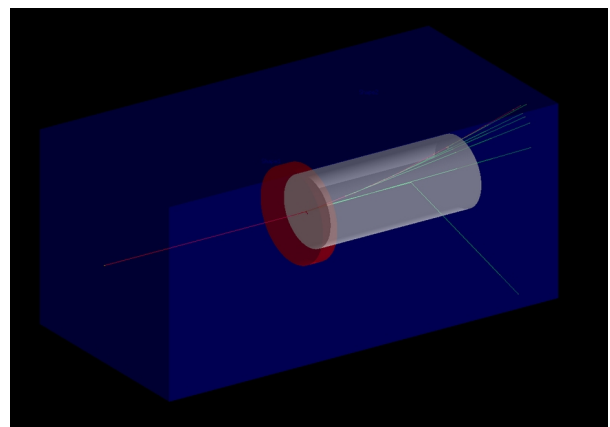

*Figure 2: Geant4 simulation model of iron doped epoxy, red is 1.5cm iron shield, white is a plastic volume to capture energy and to serve as an indicator in punch-through events.*

*3.2 Experimental*

The second portion of this study was to understand our ability to create a prototype of the shielding and that it not only meets the NASA re-entry requirements, but also the densities proposed in the simulations. This was accomplished by performing a series of experiment to ensure consistence in the manufacturing process along with repeatability. For these initial test only iron filler was used due to the cost of tungsten filler, then the prototype were fabricated out of both materials to determine if the methods were repeatable and achieved the same level of precision.

### 3.2.1 Mixing Ratio

To start we made small rectangular samples of both the base epoxy and iron doped epoxy to analyze. Calculations were made based on weight to ensure we were getting the correct epoxy to iron ratio that would result in the correct density desired. The epoxy is a two part resin that is mixed to equal amounts. For this test the total weight of epoxy was 10g to test the ease of mixing of the iron filler and to test the pouring capabilities. These samples were mixed in small cylindrical plastic containers that were cut away after full cure. The iron filler was added based on mass to give a ratio of parts iron/1 part epoxy. The cure time for the epoxy is 7 day work and 30 day full hardened, so to speed up this process the samples were placed with in a heating closet at a temperature of approximately 110°F. Once the samples were created, hardened, and cut into rectangular samples (Figure 3), we calculated the density by using a water displacement test using a graduated cylinder.

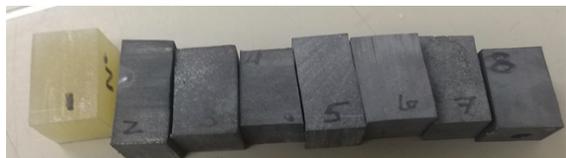

*Figure 3: Picture of small iron doped epoxy samples increasing in densities with 1 being base epoxy and 8 being the highest density .*

### 3.2.2 Height Density

The next area we studied was how the density of the iron doped epoxy would change with a factor of height. To accomplish this, we created iron doped epoxy at three separate densities with the target ratios at approximately 1:4, 1:4.5, and 1:5.5 respectively and then filled graduated cylinders. We then cut the doped epoxy into small sections with corresponding numbering (with 1 being the base of the graduated cylinder) and tested each section individually (Figure 4). These test results are discussed in section 4.2.2, but they shows that the expected results and experimental result differed significantly. We concluded that the discrepancies in the densities were attributed to the presence of air bubbles in the iron doped epoxy.

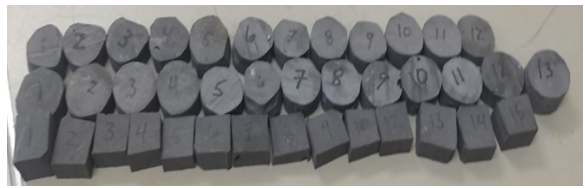

*Figure 4: Picture of sectioned graduated cylinder test (Top Row- 1:3.765 ratio, Middle Row- 1:4.619 ratio, Bottom Row- 1:5.425 ratio)*

### 3.2.3 Prototypes

For the final phase of the density study, we created several prototype shields. The first 3 shield prototypes consisted of a 1:8 ratio of epoxy to iron, with the only significant differences being the method in which we filled the molds along with the mold material itself. This was to ensure that we understood the fabrication and the techniques that are required to produce precision results. In the first prototype the iron doped epoxy was mixed in a metal container and poured into a thin cylindrical PETG mold. As expected, the density was less than what was calculated due to the formation of air bubbles. It proved to be difficult to remove due to adhesion to the PETG. The second prototype that we constructed was done so by compacting the iron doped epoxy by hand within a glass container to try and reduce air pocket. This shield has fewer air bubbles then the first but were still visible once the shield was removed from the container. This also proved difficult to extract from the mold and the shield needed to be machined out. The third shield prototype was made using a 3D mold (Figure 5), in the hopes that it would be easier to break free. During the iron doped epoxy pouring stage, we filled the mold a third of the way and using a negative of the mold, compacted it and then used a vacuum chamber to release the air bubbles. This step was repeated three times until the mold was full, with the low vacuum being applied until the epoxy reached the top of the mold then quickly releasing the pressure to cause compression at each of the three changes. This method showed a significant decrease in the number of air bubbles. A small portion of all three shields were used to determine if the density had changed with gravity as in the prior test, which held true for all prototypes.

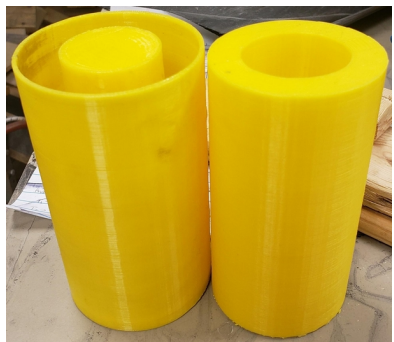

*Figure 5: 3D printed shield prototype mold and negative*

After the success of the iron doped epoxy shields, the WSU team was confident to test the process with tungsten filler. A 3D printed mold of the correct size and dimensions was created, and low vacuum techniques were repeated. Since less filler was needed to achieve the same density, creating this prototype proved easier. Issues still arose when trying to remove the hardened shield from the plastic printed mold. This led the team at WSU to try and find a new material that would allow us to easily accomplish this. The team chose to use a silicon base mold that not only allows us to remove the shield with little effort, but it also allows for reuse. The final mold (Figures 6 and 7) was made of this material and included not only the exact parameters of the shield, but also parts of the shield attachment assembly as discussed in section 3.3.

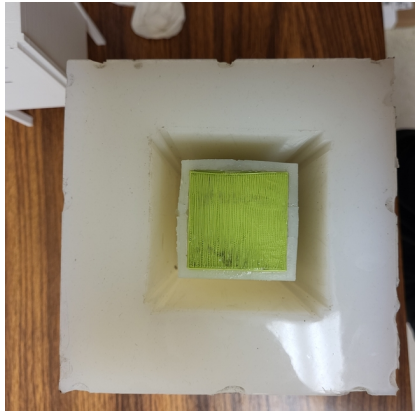

*Figure 6: Silicone shield mold with assembly attachment components*

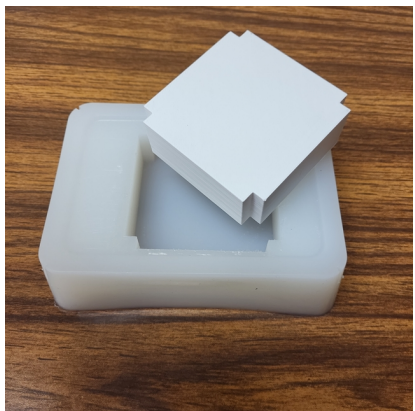

*Figure 7: Silicone shield end-cap mold*

*3.3 Computer Aided Design and Shield Attachment Assembly*

Now that a shield was designed, tested, and fabricated, a way to attach it to the CubeSat bus was needed. This was achieved by creating a CAD model of the shield and attachment assembly (Figure 8) to ensure that it could securely fit within the CubeSat bus (Figure 9). The design consists of three major components, angle brackets, axle keys, and metal plates. The angle brackets are a custom design that attach directly to the shield itself. This is accomplished by inserting the brackets into the silicone mold and pouring the doped epoxy, creating a strong bond between the materials. The metal plates are mounted on opposites sides and attach to the CubeSat bus directly through tapped holes. The fabricated design is intended to decrease the weight while still being structurally sound. By doing this the plates prohibit movement and provide lateral strength to the shield. To prevent axial movement and to secure the shield, axle keys are added in between the angle brackets and the metal plates. Two of these keys will be joined by a layer of epoxy and all four will be attached through tapped bolts. The final fabrication of the tungsten doped epoxy shield and attachment assembly started in September 2022 and will be completed by December 2022 with the full CubeSat construction being completed in fall of 2023.

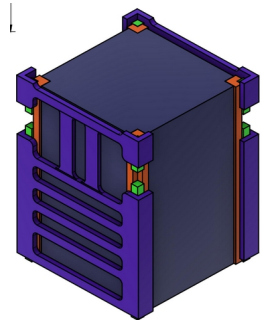

*Figure 8: Isometric Shield Attachment CAD Model*

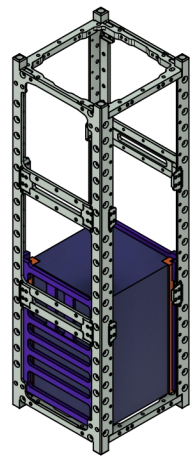

*Figure 9: Isometric CubeSat and Shield CAD Model*

**4. Results and Discussion**

By recognizing the need to protect the science package from Cosmic Rays, the nuSOL team was able to simulate, design, and test a new innovative type of radiation shield. The results of this study demonstrate that an effective shield is achievable and gives the advancement of scientific research not only to the nuSOL project but to future projects as well.

*4.1 Simulation results*

During this phase of the study, WSU team conducted separate simulations. The first consisted of shooting both electrons and protons at energies ranging from 1MeV to 1GeV. These particles were fired at a 1.5cm thick shield that consisted of iron doped epoxy ranging in densities from $2g/cm^3$ to $5g/cm^3$ or tungsten doped epoxy ranging from $3.5g/cm^3$ to $7.5g/cm^3$. This simulation was to determine the characteristics of the doped epoxy and to analyze the radiation shielding capabilities when compared to the standards of the base epoxy and solid steel. At the lowest simulated density the iron doped epoxy has the ability to completely stop 6MeV electrons and 52MeV protons. Also at this density a 90% punch-through rate did not occur until 14MeV and 54MeV for electrons and protons, respectively. The upper limit on iron doping has the ability to completely stop electrons and protons at 12MeV electrons and 94MeV protons. With a 90% punch-through rate of 28MeV electrons and 102MeV protons. For tungsten doped epoxy the results increased significantly when compared to the iron doped epoxy. With the lower density having a complete stopping power up to 12MeV electrons and 60MeV protons, and a 90% punch-through rate of 74MeV and 64MeV respectively. On the higher density of $7.5g/cm^3$ the complete stopping power resulted in 20MeV electrons and 85MeV protons. With a 90% punch-through rate of 1GeV electrons and 90 MeV protons. While the iron doped epoxy at the density of $5g/cm^3$ does perform better than the tungsten for higher energy protons, as discussed in section 4.2, the highest density for an iron doped epoxy shield to be structurally achievable is $4g/cm^3$. The consisted of one million individual particle events and resulted in a statistical error of less than 0.1% at the 5-sigma level.

As expected, the simulations show an increase in stopping power as we trend towards the density of steel with the differences decreasing between each increase in density. Given thw weight requirements for the 3U Cube-sat this gave us an understanding for the optimization needed to meet not only the weight requirement but also the shielding capabilities. These results when compared to the standard of a solid steel plate (Figures 10 and 11) show that not only does the tungsten shield out-perform the iron shield, but also out-preforms the solid steel shield in regards to electrons. These simulations show that it is possible to have some Cosmic Ray radiation shielding, with the nuSol team considering a 10% reduction in Cosmic Ray noise a success, and that meets weight and re-entry requirements.

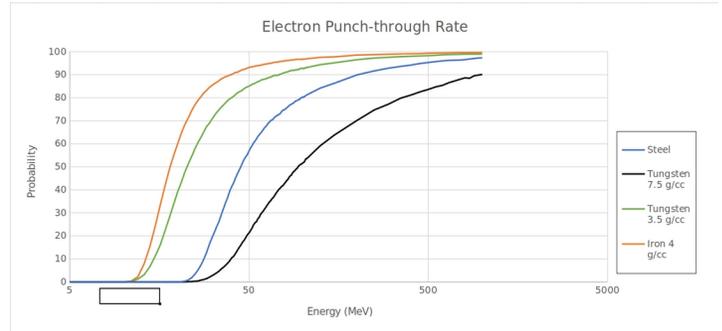

*Figure 10: Graph of Simulated electron Punch-through Rates*

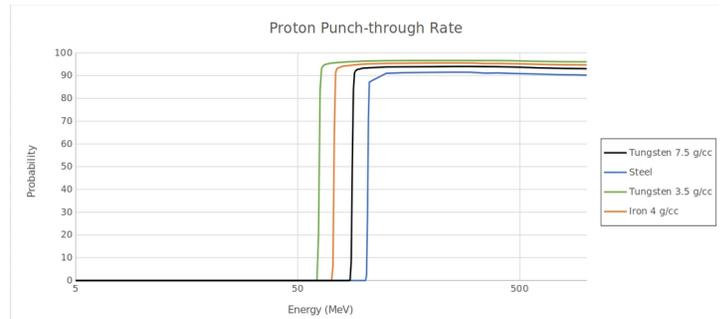

*Figure 11: Graph of Simulated proton Punch-through Rates*

The second portion of the simulations was to see how the final $7.5g/cm^3$ tungsten doped epoxy shield would handle other particles such as alpha, and oxygen and iron nuclei. This study shows that at high energies the alpha particles tend to plateau at approximately 90% punch-through rate and that the nuclei never exceed the 10% noise reduction point(Figure 12). Indicating that even at high energies the tungsten doped epoxy shield is a success.

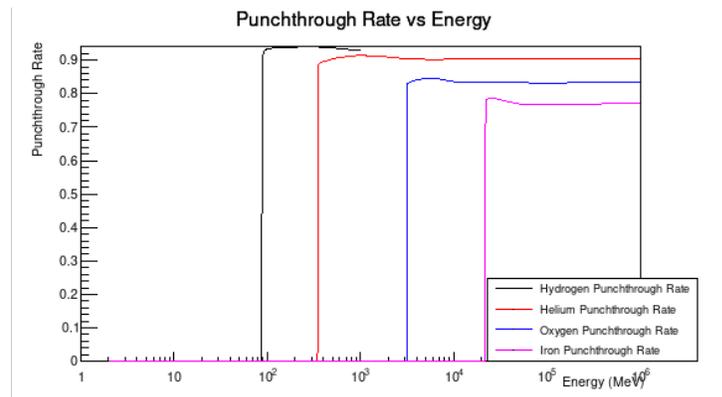

*Figure 12: alpha Particle, and oxygen and iron nuclei Punch-through rates*

### 4.2 Experimental Results

After completion of simulations, the WSU team needed to ensure that actual construction of a shield was achievable. In following the methods outline in section 3, we were able to create test samples of the iron doped epoxy to have physical evidence of the fabrication and manufacturing process. This resulted eventual manufacturing of both an iron and tungsten doped epoxy shield prototypes. This was to understand if the simulated densities were possible and to also test the structural capabilities of the shield itself. While an iron doped epoxy of $5g/cm^3$ simulated performs best, the mixture becomes supersaturated and never fully hardens making it fail structurally. As higher densities of iron were tested, the mixture becomes concrete in nature and proves harder to pour and manipulate. The highest achievable mixture results in a density of $4g/cm^3$ for iron. The issue does not arise with tungsten doped epoxy at tested densities. This is due to the weight limits of the CubeSat, so further samples were not tested beyond that limit.

#### 4.2.1 Mixing Ratio

The initial step to understand, was how the iron filler would mix with the epoxy and how easily it could be manipulated. To accomplish this the WSU team, created small samples to test if the calculated densities coincided with real life manufacturing. Calculations were made based on weight to ensure we were getting the correct epoxy to iron ratio that would result in the correct density desired. Once the samples were created, we calculated the density by using a water displacement test using a graduated cylinder. The results show that the lower densities are more easily achieved, the higher ones had a significant error (Figure 13). This is due to the creation of air bubbles during the mixing process and later indicated that low vacuum techniques were required.

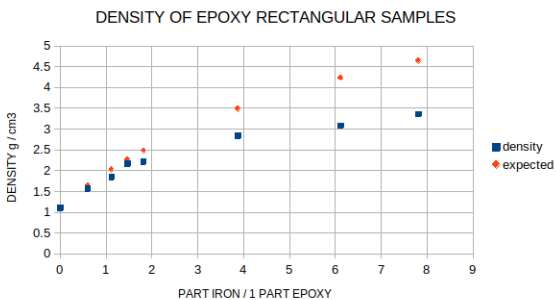

*Figure 13: Graph of Expected and Obtained iron Doped Epoxy in Small Samples*

#### 4.2.2 Height Density Study

The next test not only included the ability to mix the iron filler with epoxy, but also test if a change in density occurred with a height. This test is important given that the final shield is 12.5cm tall and if the density was not homogeneous the shield would be ineffective and create Cosmic Ray noise discrepancies based on positioning of the CubeSat. After full cure the samples were cut into smaller sample sizes and the densities were tested using water displacement. With the sections not being completely uniformed and the error of 10% between the scale used to weigh the sections and the graduated cylinder, we concluded the density was not changing with height significantly (Table 1). This resulted in average overall densities resulting in 3.133 $g/cm^3$ for a 1:3.765 ratio, 3.3383 $g/cm^3$ for a 1:4.619 ratio, and 3.183 $g/cm^3$ for a 1:5.425 ratio. As mentioned previously mixing accurately was still a concern, and as before the lower densities proved more accurate then higher samples (Figure 14). This indicated that the formation of air bubbles was still an issue.

| CUT # | WEIGHT (g) | DISPLACEMENT (ml) | Density (gm/cm3) | IRON PARTS PER 1 PART EPOXY |
|---|---|---|---|---|
| 1 | 7.7 | 2.5 | 3.08 | 3.765 |
| 2 | 8 | 2.75 | 2.91 | |
| 3 | 7.8 | 2.5 | 3.12 | |
| 4 | 7.9 | 2.75 | 2.87 | |
| 5 | 7.8 | 2.5 | 3.12 | |
| 6 | 7.2 | 2.25 | 3.2 | |
| 7 | 5.8 | 1.75 | 3.31 | |
| 8 | 6.9 | 2.25 | 3.07 | |
| 9 | 6.5 | 2 | 3.25 | |
| 10 | 7.1 | 2.25 | 3.16 | |
| 11 | 7.4 | 2.5 | 2.96 | |
| 12 | 3.9 | 1.1 | 3.55 | |

*Table 1: Sample Data Table for Graduated Cylinder Test (3.765 Parts iron to 1 Part Epoxy)*

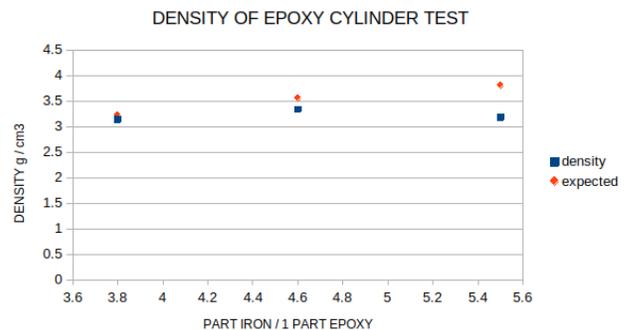

*Figure 14: Graph of Expected and Obtained iron Doped Epoxy in Height Density Samples*

#### 4.2.3 Prototypes

After completing the initial experiments, the nuSOL team created a prototype of the radiation shield. During this process four prototypes were created, three made of iron doped epoxy and one tungsten doped epoxy. The first iron prototype was created in the same manner of the previous samples. With an expected density of $3.6g/cm^3$ and an actual of 2.8. this indicated to the WSU team that low vacuum techniques needed to be tested to reduce the formation of air bubbles. These techniques were used on the following two iron doped epoxy prototypes. By doing so the expected densities and actual lined up more accurately (Figure 15). After studying the iron doped epoxy prototypes the WSU team was confident in their ability to create a precision shield, a tungsten doped epoxy shield was made with correct dimensions (Figure 16). This resulted in a structurally strong shield with a density of 3.48g/cm$^3$ and has an accuracy of 99.4% of the expected value of 3.5g/cm$^3$. the last requirement that needed to be tested was a what would happen when the shield re-entered Earth's atmosphere and if any piece would hit with more than 15J of energy. To do this, small samples were made during prototype pouring process, hardened, and burned with a torch. The iron doped epoxy sample started burning at approximately 260°C, with full destruction at 700°C. The tungsten doped epoxy sample started burning at 315°C and was destroyed at 424°C. Both of this are well below the average re-entry temperature of 1800°C.

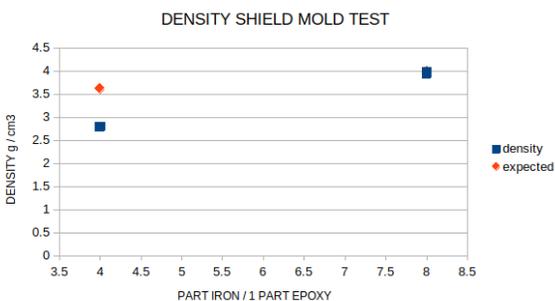

*Figure 15: Graph of Expected and Obtained iron Doped Epoxy in Prototypes*

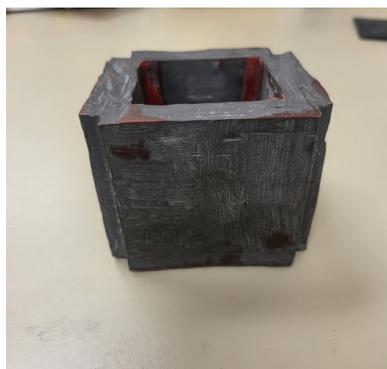

*Figure 16: Picture of tungsten Doped Epoxy Prototype*

### 5. Conclusions

The nuSol project 3U CubeSat mission is vitally important to the further advancement in the study of neutrinos. If the CubeSat mission is successful, then a satellite mission orbiting close to the Sun will be within reach. To ensure the success of the CubeSat, new and innovative technologies are required. One such obstacle was to create a Cosmic Rays radiation shield to reduce the potential of noise within the science package. The team at WSU was tasked with accomplishing this feat. By not only understand the science behind the mission, but the engineering, the team was able to accomplish this, through simulations, experimental data, and finally working prototypes. Further testing is scheduled to ensure the shield performs as expected, but the ability to stop or reduce Cosmic Rays radiation noise with several never reaching the 90% punch-through rates is ground-breaking for the science community. Not only does the shield meet all requirements for the CubeSat mission, but also that of NASA. This will be the first mission flown with this type of shielding and by showing that it is possible, the advancement of capabilities of other missions will greatly enhance, furthering scientific research

### Acknowledgments

I would like to thank Dr. Nick Solomey for leading the nuSOL project and giving me the opportunity not only to excel in physics but also in engineering. I would also like to thank PhD candidate Jonathan Folkerts for helping lead the research team at WSU and for his immense contributions to the nuSOL project and for spearheading the research for the Sun orbiting satellite. Thanks to Trent English for his contributions that led to the success of the shield design.

### References

[1] N. Solomey, J.Folkerts, H. Meyer, C. Gimar, J. Novak, B. Doty, T. English, L. Buchele, A. Nelsen, R. McTaggart, M. Christl, Design of a Space-based Near-Solar Neutrino Detector for nuSOL experiment, Arxiv (2022) https://doi.org/10.48550/arXiv.220600703

[2] N.solomey, A. Dutta, H. Meyer, J. Folkerts, T. English, K. Messick, J. Novak, J. Chee, M. Christl, R. McTaggart, and et al., Niac phase-2 final report for astrophysics and technical lab studies of a solar neutrino spacecraft detector, NASA Technical Report Server, Apr 2022

[3] 3M™ Scotch-Weld™ Epoxy Adhesive 2216 B/A Translucent Material Properties https://www.alliedelec.com/m/d/badcaf2fd3816d3706fc411c0bf6ab8c.pdf#page=6&zoom=auto,-267,559/, (accessed 20.07.2021)